\documentclass[twocolumn,showpacs,preprintnumbers,amsmath,amssymb]{revtex4}

\usepackage{graphicx}
\usepackage{dcolumn}
\usepackage{bm}

\begin{document}


\title{Gate-controlled nuclear magnetic resonance \\in an AlGaAs/GaAs quantum Hall device}

\author{S. Masubuchi, K. Hamaya, and T. Machida\footnote{electronic address : tmachida@iis.u-tokyo.ac.jp}}%
\affiliation{Institute of Industrial Science, The University of Tokyo,  \\4-6-1 Komaba, Meguro-ku, Tokyo 153-8505, Japan.
 }%


\date{\today}

\begin{abstract}
We study the resistively detected nuclear magnetic resonance (NMR) in an AlGaAs/GaAs quantum Hall device with a side gate. The strength of the hyperfine interaction between electron and nuclear spins is modulated by tuning a position of the two-dimensional electron systems with respect to the polarized nuclear spins using the side-gate voltages. The NMR frequency is systematically controlled by the gate-tuned technique in a semiconductor device. 

\end{abstract}

\maketitle

For the processing of quantum information with multiple quantum bits, ideal quantum states with various discrete energies have been required in solid states. In a proposal of a silicon-based nuclear spin quantum computer,\cite{Kane} a concept of the gate-tuned hyperfine interaction between electron and nuclear spins was presented in order to individually address the nuclear spins. To implement this concept in electron spin systems, the gate control of the electron spin resonance (ESR) in Si/SiGe heterostructures was also suggested:\cite{Vrijen} by employing materials with different $g$-factors, the electrical gating can control the ESR frequency, based on a displacement of the electron wavefunction within the heterostructures. Recently, Jiang {\it et al.}\cite{Jiang} clearly demonstrated the controlled ESR frequency of the confined two-dimensional electron systems (2DES) in AlGaAs/GaAs heterostructures with the direct $g$-factor control by means of tuning the gate voltage.   

To date, Poggio {\it et al.}\cite{Poggio} employed the displacement of the electron wavefunction by external gate voltages in AlGaAs parabolic quantum wells (QWs), which resulted in a local manipulation of the position of the nuclear spin polarization. Also, Sanada {\it et al.}\cite{Sanada} used gate-induced changes in the spin relaxation time of electrons in GaAs(110) QWs,\cite{Sanada} causing an enhancement in the dynamic nuclear polarization (DNP). In fractional quantum Hall (QH) systems, the nuclear spin relaxation time was changed by the gate voltage using the resistive detection of the DNP.\cite{Hashimoto,Smet,Yusa} However, no evident demonstration of the gate-controlled NMR has been reported yet, because it seems to be difficult to tune the strength of the interaction between electrons and nuclei and to detect the systematic change in the NMR frequency.
\begin{figure}[b]
\includegraphics[width=7cm]{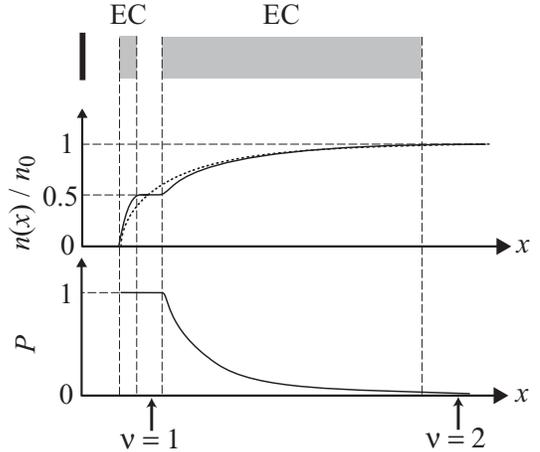}
\caption{[Top] A top view of the edge channels (EC): the electron system is separated into compressible (gray) and incompressible (white) strips in the presence of screening effect.\cite{Chklovskii} [Middle] Diagrams of the local electron density near the QH edge channels with (solid curves) and without (dashed curves) an external magnetic field. [Bottom] The local electron spin polarization ($P$). }
\end{figure} 

In this letter, we report on the first experimental demonstration of the gate-controlled NMR frequency ($f_\mathrm{NMR}$) in a semiconductor heterostructure. A method for the detection of the Knight shift ($K_\mathrm{S}$) was recently developed using a resistively detected NMR (RDNMR) technique in an integer QH device with a side gate (SG):\cite{Masubuchi} the removal of the electron system near the polarized nuclear spins with the SG voltages ($V$$_\mathrm{SG}$) enabled us to turn off the hyperfine interaction. By applying this method to tuning the strength of the hyperfine interaction, we evidently show a systematic modulation of $f_\mathrm{NMR}$ by the $V$$_\mathrm{SG}$.
\begin{figure}[t]
\includegraphics[width=7cm]{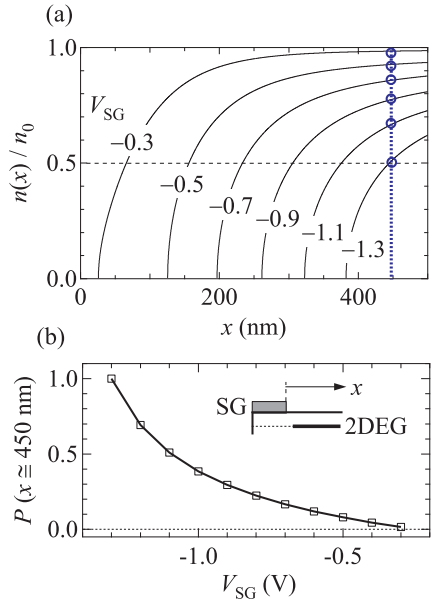}
\caption{(a) The calculated $n$($x$)/$n_\mathrm{0}$ vs $x$ for $V_\mathrm{SG} =$ -0.3, -0.5, -0.7, -0.9, -1.1, and -1.3 V. The DNP is formed in the region around $n$($x$)/$n_\mathrm{0} =$ 0.5. When the nuclear spins are polarized at $V_\mathrm{SG} =$ -1.3 V, the DNP is formed at $x \sim$ 450 nm (blue line). The 2DES depth from the surface is $d =$100 nm and the pinch-off voltage is - 0.24 V. (b) The $P$ calculated from Eq. (1) at $x \sim$ 450 nm for various $V_\mathrm{SG}$. The inset shows the cross-sectional illustration of the 2DES near the SG.}
\end{figure} 

The following is the concept of our experiment for modulating $f_\mathrm{NMR}$ by $V$$_\mathrm{SG}$. As shown in Fig. 1, electron spin polarization is $P =$ 0 in the bulk region of the Landau level filling factor $\nu =$ 2, because both the levels with spin-up and spin-down are filled. In approaching to the edge region, the local electron density $n$($x$)/$n_\mathrm{0}$ is reduced owing to the confinement potential at the sample edge, where $x$, $n$($x$), and $n_\mathrm{0}$ are the distance from the SG edge toward the inner region [inset of Fig. 2(b)], the local electron density at $x$, and the electron density in the bulk region, respectively. In the presence of the screening effect, the electron system is separated into compressible and incompressible strips\cite{Chklovskii} as depicted in the top illustration of Fig. 1.  
The width of the compressible region between $\nu =$ 1 and 2 is about 1 $\mu$m,\cite{Oto} which is sufficiently larger than the width of the DNP ($\sim$10 nm). Thus, the $P$ continuously changes from 1 to 0 over macroscopic region. Since it is well known that the $K_\mathrm{S}$ is proportional to $P$ and $n$,\cite{Slichter} we can utilize the control of the position of the edge channels by the $V$$_\mathrm{SG}$ as the modulation technique to tune the strength of the hyperfine interaction.
\begin{figure}[t]
\includegraphics[width=8.5cm]{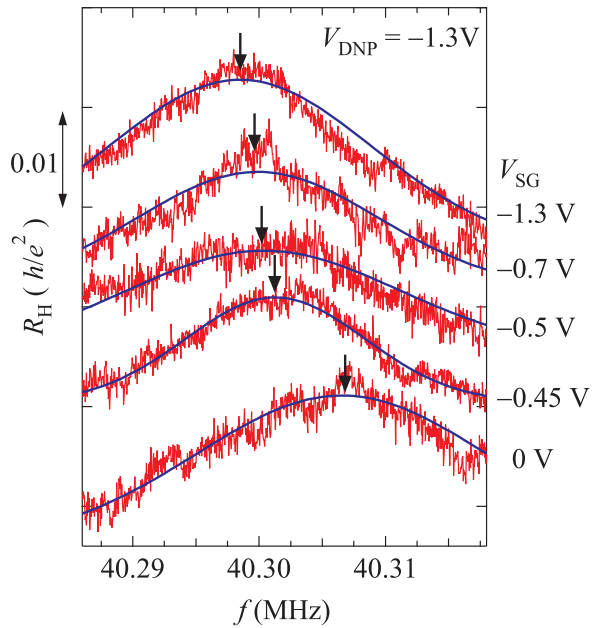}
\caption{The NMR spectra detected by the change in the Hall resistance for $^{71}$Ga nuclei. The DNP is formed at $V$$_\mathrm{SG} = V$$_\mathrm{DNP} =$ -1.3 V and the $B_\mathrm{RF}$ is applied at various $V_\mathrm{SG}$. The blue lines show the Gaussian curves. }
\end{figure}

The Al$_{0.3}$Ga$_{0.7}$As/GaAs QH device in this study is $K$-shape Hall bar geometry \cite{Machida} with two cross gates and a SG used in our previous work.\cite{Masubuchi} The electron density and mobility of the sample are 1.5 $\times$ 10$^{15}$ m$^{-2}$ and 40 m$^{2}$/Vs, respectively. When the filling factor in the bulk region is $\nu_\mathrm{B} =$ 2 ($B =$ 3.1 T), two QH edge channels are formed along the edge regions of the Hall bar sample. Adjusting the filling factor beneath the two cross-gates to $\nu_\mathrm{G}$ = 1, we selectively populate spin-resolved edge channels at the position between the cross gates, and the nuclear spins are polarized in the narrow region ($\sim$ 10 nm) between the relevant edge channels.\cite{Machida} The DNP and its depolarization were detected by measuring the Hall resistance ($R_\mathrm{H}$).\cite{Masubuchi,Machida} Transport measurements were carried out by standard ac method ($I$$_\mathrm{AC}$ $=$ 1.0 nA) in a $^{3}$He-$^{4}$He dilution refrigerator at 50 mK. To make the DNP between edge channels, we applied $I$$_\mathrm{DC} =$ +7 nA to $I$$_\mathrm{AC}$. In the experimental procedure, we first formed the DNP at a specific position for $V$$_\mathrm{SG}$ = -1.3 V. Secondly, to tune the position of the 2DES, we changed a $V$$_\mathrm{SG}$, and then the rf magnetic fields ($B_\mathrm{RF}$) are applied for 2 s. Next, the $V$$_\mathrm{SG}$ was returned to the value of -1.3 V in order to record the $R_\mathrm{H}$. Finally, we plotted $R_\mathrm{H}$ as a function of rf frequency (NMR spectrum) for various $V$$_\mathrm{SG}$ values. The detailed procedure for the similar measurements was reported in Ref. 9.

Using the calculation given by Larkin and Davies,\cite{Larkin} we can obtain the electron density profiles, $n$($x$)/$n_\mathrm{0}$ vs $x$, for various $V_\mathrm{SG}$ [Fig. 2(a)]. Since the DNP is formed in the $\nu =$ 1 region,\cite{Masubuchi,Machida} the center of the DNP region corresponds to $n$($x$)/$n_\mathrm{0} =$ 0.5 at $B =$ 0 T (Fig. 1). Here, the position of the DNP formed at $V_\mathrm{SG} = V_\mathrm{DNP} =$ -1.3 V is $x \sim$ 450 nm, estimated from the calculated results in Fig. 2(a). After the DNP at $x \sim$ 450 nm, we vary the $V_\mathrm{SG}$ and apply $B_\mathrm{RF}$, and then the DNP interacts with the 2DES with various $n$($x$)/$n_\mathrm{0}$ [open circles in Fig. 2(a)]. 
When we focus on the region from $\nu =$ 2 to 1 [$n$($x$)/$n_\mathrm{0}$ $\geq$ 0.5], the $P$ can be written as
\begin{equation}
P = \frac{1}{n(x)/n_\mathrm{0}} - 1. 
\end{equation}
Accordingly, the $P$ is reduced gradually with varying $V_\mathrm{SG}$ at this position [Fig. 2(b)]. As the DNP interacts with the 2DES with the calculated $P$, $f_\mathrm{NMR}$ $vs$ $V_\mathrm{SG}$ is  estimated as shown in inset of Fig. 4 using NMR frequency of GaAs substrate ($f_\mathrm{GaAs}$) and the maximum Knight shift value ($K_\mathrm{S}^{P=1}$).
\begin{figure}[t]
\includegraphics[width=8cm]{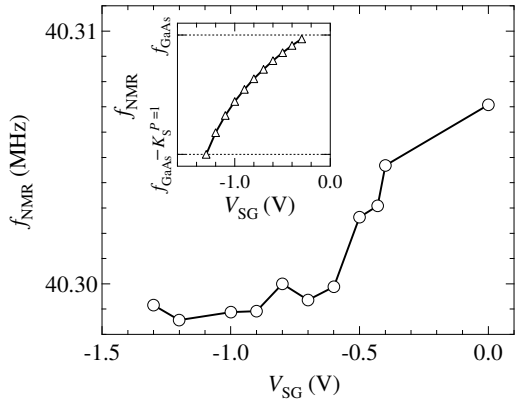}
\caption{The plot of $f_\mathrm{NMR}$ as a function of $V_\mathrm{SG}$. The inset shows the calculated $f_\mathrm{NMR}$.}
\end{figure}
 
Figure 3 shows a sequence of NMR spectra for $^{71}$Ga nuclei for various $V_\mathrm{SG}$ values. Here, the nuclear spins are dynamically polarized at $V_\mathrm{DNP} =$ -1.3 V ($x \sim$ 450 nm), and the $B_\mathrm{RF}$ is applied after the position of the 2DES is changed by each $V_\mathrm{SG}$. With increasing $V_\mathrm{SG}$, that is, with reducing the hyperfine interaction, the $f_\mathrm{NMR}$ is evidently shifted from 40.2985 to 40.3065 MHz. For the measurement at $V_\mathrm{SG} =$ 0 V, the position of the DNP exists in $\nu =$ 2 regime, so that the DNP interacts with the 2DES with $P =$ 0: the $K_\mathrm{S}$ is nearly equal to 0. 

We summarize the $f_\mathrm{NMR}$ as a function of $V_\mathrm{SG}$ in Fig. 4. The $f_\mathrm{NMR}$ is systematically changed by adjusting the $V_\mathrm{SG}$. We can see that the experimental trend of the increase in the $f_\mathrm{NMR}$ with varying $V_\mathrm{SG}$ is qualitatively similar to that shown in the inset of Fig. 4. It is speculated that the shape difference between the experimental result and the calculation is attributed to the presences of the screening effect in the edge-channel dispersion\cite{Chklovskii} and the local distribution of the polarized nuclear spins. Although the controlled quantum states are arising from an ensemble consisting of $\sim$10$^{9}$ nuclear spins, the present result is the first experimental demonstration of gate-controlled NMR in semiconductors. Since the present technique utilizes the advantage of semiconductor devices such as the gate controllability of electron density and distribution of the electron spin polarization, we hope that this technique can become key to achieve a processing of a nuclear spin-based quantum information. 
 
In summary, we have experimentally demonstrated the gate-controlled NMR in a semiconductor-based QH device. To tune the strength of the hyperfine interaction between electron and nuclear spins, the positions of the 2DES with respect to the polarized nuclear spins are changed by the side-gate voltages. The systematic change in the NMR frequency is clearly shown. 

This work is supported by PRESTO, JST Agency, the Grant-in-Aid from MEXT (No. 17244120), and the Special Coordination Funds for Promoting Science and Technology. K. H. acknowledges JSPS Research Fellowships for Young Scientists. 


\end{document}